\documentclass[aps,twocolumn,prl,superscript,floatfix,superscriptaddress,10pt]{revtex4-2}

\usepackage{epsfig}
\usepackage{graphicx}
\usepackage{dcolumn}
\usepackage{bm}
\usepackage{amsfonts}
\usepackage{amsmath}
\usepackage{amssymb}
\usepackage{mathrsfs}
\usepackage{natbib}
\usepackage{multirow}
\usepackage{lineno}

\usepackage[colorlinks=true,%
            linkcolor=blue,%
            urlcolor=blue,%
            citecolor=blue,%
            filecolor=blue,%
            bookmarksopen=true,%
            pdfauthor={M.Figueroa},%
            pdftitle={QR},%
            pdfsubject={QR_manuscript},%
            pdfpagemode=UseOutlines]{hyperref}

\usepackage[T1]{fontenc} 
\usepackage{lmodern}
\usepackage{xcolor}

\begin{document}

\title{Exceptional flat bands in bipartite non-Hermitian lattices}
\author{Juan Pablo Esparza}
\affiliation{Departamento de F\'isica, Universidad T\'ecnica Federico Santa Mar\'ia, Casilla 110, Valpara\'iso, Chile}
\affiliation{Instituto de F\'isica, Pontificia Universidad Cat\'olica de Valpara\'iso, Avenida Universidad 331, Curauma, Valpara\'iso, Chile}

\author{Vladimir Juri\v ci\'c}\thanks{Corresponding author:vladimir.juricic@usm.cl}
\affiliation{Departamento de F\'isica, Universidad T\'ecnica Federico Santa Mar\'ia, Casilla 110, Valpara\'iso, Chile}

\begin{abstract}
Flat bands, in which kinetic energy is quenched and quantum states become macroscopically degenerate, host a rich variety of correlated and topological phases, from unconventional superconductors to fractional Chern insulators. In Hermitian lattices, their formation mechanisms are now well understood, but whether such states persist, and acquire new features in non-Hermitian (NH) { crystals}, relevant to open and driven systems, has remained an open question. Here we show that the Hermitian principle for flat-band formation in bipartite lattices, based on a sublattice degeneracy mismatch, extends directly to the NH regime: whenever one sublattice hosts a momentum-independent eigenvalue with degeneracy exceeding that of its partner on the other sublattice, flat bands arise regardless of gain, loss, or complex couplings. Strikingly, at  exceptional points, dispersive bands coalesce to form \emph{exceptional flat bands} (EFBs) that persist beyond these singularities, exhibiting biorthogonal eigenmodes spanning both sublattices, with energies and lifetimes tunable via sublattice asymmetry and non-reciprocal couplings. This general framework unifies Hermitian and NH flat-band constructions, and reveals dispersionless states with no closed-system analogue, as is the case of a bipartite lattice with imbalanced but constant sublattice chemical potentials. The proposed construction is applicable to synthetic platforms, from classical metamaterials, where flat bands can be directly emulated, to quantum-engineered systems such as photonic crystals and ultracold atom arrays, which should host correlated and topological phases emerging from such EFBs.
\end{abstract}

\maketitle


Flat electronic bands, where kinetic energy is completely  quenched and quantum states exhibit a macroscopic degeneracy, are fertile ground for unconventional superconductivity, fractional Chern insulators, and other exotic many-body phases~\cite{Bergholtz2013fractional,Parameswaran2013Fractional,heikkila2016flat,volovik2018graphite,Roy2019Unconventional,cao2018correlated,cao2018unconventional,Cao2021Nematicity,xie2021fractional,park2021flavour,Das2021Symmetry,Choi2021Correlation,Sanchez2024Nematic}. 
The theoretical proposals~\cite{suarez2010flat,bistritzer2011moire,dosSantos2007}  and experimental  discovery~\cite{cao2018correlated,cao2018unconventional} of nearly dispersionless bands in twisted bilayer graphene (TBG) revealed the powerful role of symmetry and topology in stabilizing such states~\cite{kang2018symmetry,po2019faithful}, sparking intense efforts to design flat-band systems across condensed matter~\cite{shores2005structurally,mendels2010quantum,han2012fractionalized,mazin2014theoretical,lisi2021observation,hase2024new}, metamaterials~\cite{Nakata2012Observation,Kajiwara2016Observation,Wang2019Highly}, photonics~\cite{Takeda2004Flat,Vicencio2015Observation,Baboux2016Bosonic,Mukherjee2017Observation}, and ultracold atomic platforms~\cite{He2017Realizing,Cooper2019Topological,Zeng2023Observation,Sui2025Topologically}. 
While the symmetry-based mechanisms behind flat bands in Hermitian crystals are now well understood~\cite{Ma2020Spin-orbit,cualuguaru2022general,Hwang2021General,kim2023general}, their fate in the fundamentally different setting of non-Hermitian (NH) quantum matter, relevant to open and out-of-equilibrium systems, remains largely unexplored~\cite{Leykam-PRB2017,Ramezani-PRA2017,Zyuzin-PRB2018,Zhang-PRA2019,Jin-PRA2019,Maimaiti-PRB2021,Ding2021non,Talkington2022Dissipation,Marques2022Generalized,Jiang2024exact,amelio2024lasing,Banerjee-PRB2024,leong2025NH}.

\begin{figure}[t]
    \centering
    \includegraphics[width=\linewidth]{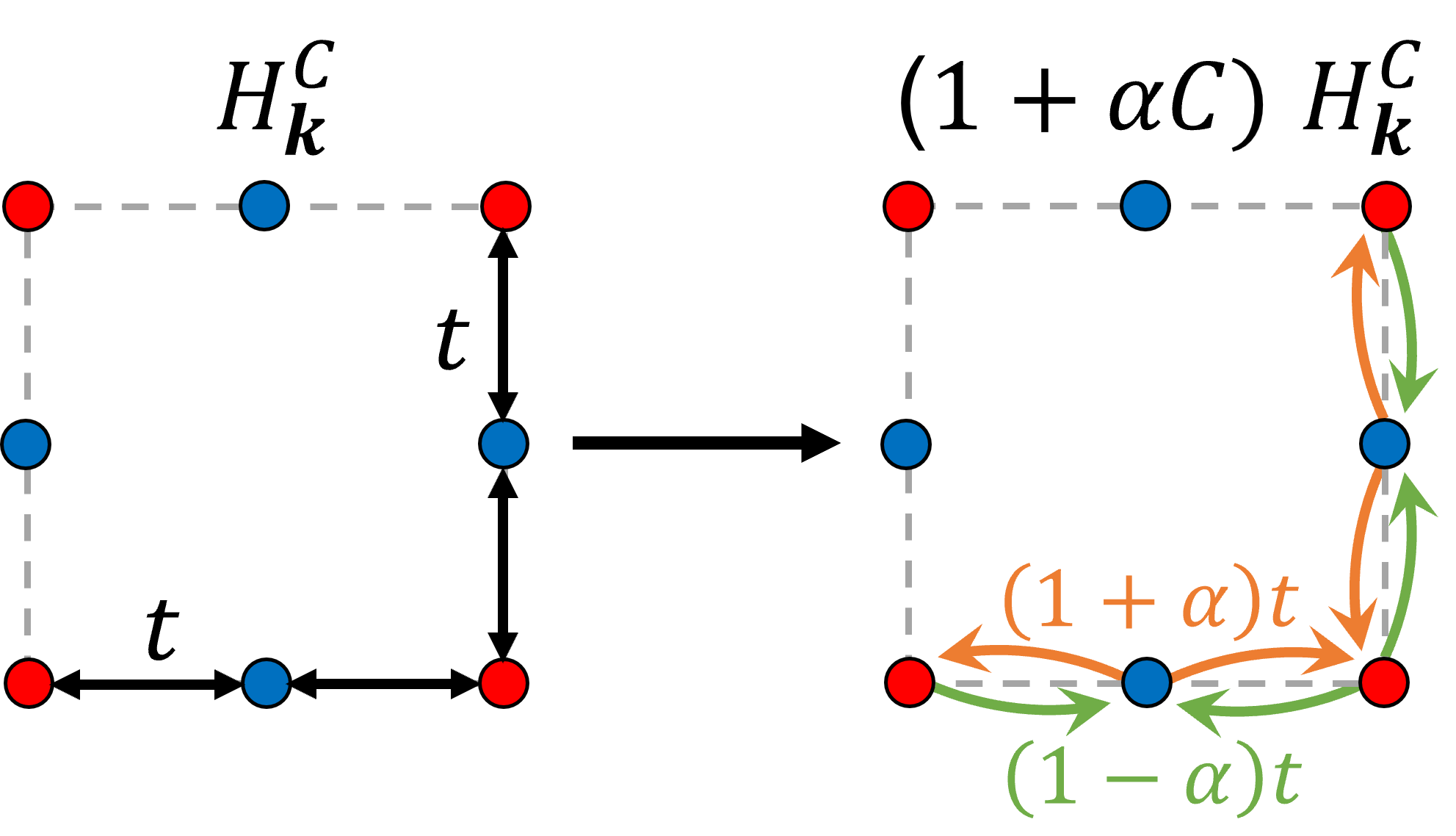}
    \caption{Construction of a non-Hermitian bipartite crystal. Left:  the square Lieb lattice with three inequivalent sites in the unit cell, forming two sublattices, $L_A$ (blue dots) and $L_B$ (red dots), with the flat band arising due to the imbalance in the number of sites belonging to each sublattice, with $N_A=2$ and $N_B=1$. Right: the non-Hermitian crystal obtained by exploiting the inherent particle-hole symmetry of the original lattice ($C$), where the new contribution acquires the form of a non-reciprocal hopping between nearest neighbors.}
    \label{fig:Sketch}
\end{figure}

NH physics, characterized by complex spectra, asymmetric couplings and engineered gain–loss, has become a central theme in synthetic and open quantum matter~\cite{Ueda-PRX2018,Foa-Torres2019perspective,Bergholtz-2021}. It departs from Hermitian band theory through two hallmark phenomena: (i) exceptional points (EPs), where eigenvalues and eigenvectors coalesce, enabling anomalous transport and topological spectral flows~\cite{Heiss_2012_exceptional,Hu-exceptional-2017,Foa-2018,yao-wang-PRL2018,yao-wang-nh-chern-2018,LeeThomale2019_PRB_AnatomySkinModes,Kozzi2024lifetime}; and (ii) the NH skin effect (NHSE), in which a macroscopic number of modes pile up at the boundary due to non-reciprocity and spectral winding~\cite{Foa-2018,yao-wang-PRL2018,yao-wang-nh-chern-2018,LeeThomale2019_PRB_AnatomySkinModes,Ueda-PRX2018,Bergholtz-2021}.
This endeavor  has recently pushed to the interacting Dirac materials~\cite{jurivcic2024yukawa,Xiang-2024,murshed2024quantum,murshed2024yukawalorentz,PinoPRB-2025}, opening a new frontier where NH effects intertwine with strong correlations.
Yet, despite reshaping our understanding of band topology, these advances have left open a central question: can flat bands, long celebrated in Hermitian systems for hosting correlated and topological phases, persist and acquire distinctive signatures in this NH landscape?

Initial insight came from NH generalizations of TBG, where it was shown that flat bands can emerge at \textit{exceptional magic angles} due to spectral collapse at NH degeneracies~\cite{esparza2024exceptional}. 
These findings suggested that non-Hermiticity could stabilize flat bands in crystalline systems under specific conditions, a possibility further supported by model-based studies across photonic and electronic platforms~\cite{Leykam-PRB2017,Ramezani-PRA2017,Jin-PRA2019,Maimaiti-PRB2021,Banerjee-PRB2024,Huang-PRB2025,leong2025NH,Marques2022Generalized}. 
However, a general and symmetry-based framework capable of predicting and engineering such flat bands has so far been lacking.

Here we show that the Hermitian principle for flat-band formation in bipartite lattices~\cite{cualuguaru2022general}, based on a sublattice degeneracy mismatch, extends directly to the NH regime: whenever one sublattice hosts a momentum-independent eigenvalue with degeneracy exceeding that of its counterpart at other sublattice, flat bands
arise irrespective of the specific NH effects, including gain, loss, or complex-valued couplings. {Furthermore,  in the NH setting, irrespective of such degeneracy mismatch,} dispersive bands can coalesce at EPs, { which can be viewed as the codimension-one   exceptional surfaces~\cite{Okugawa2019Topological,BCKB-PRB2019,Zhou2019Exceptional,
Zhou2019Exceptional}}, and persist as dispersionless  modes beyond these spectral singularities, yielding \emph{exceptional flat bands} (EFBs) with biorthogonal eigenmodes spanning both sublattices. Their energy and lifetime can be continuously tuned via sublattice asymmetries, and intra-sublattice hoppings, with EFBs emerging even for  a bipartite lattice with imbalanced but constant sublattice chemical potentials.  We showcase its versatility through examples from NH Lieb lattices, showing a geometric imprint of the EFBs on the quantum metric (Fig.~\ref{fig:quantum_metric}), to realistic multi-band models (Figs.~\ref{fig:chiral_tenband}-\ref{fig:Ca2Ta2O7}), highlighting its applicability across  diverse synthetic platforms. Because it depends only on sublattice structure, the principle is directly relevant to photonic crystals with engineered gain–loss profiles~\cite{Jin2017Topological,zhou2018optical,pan2018photonic,Ueda-PRX2018,Song2019Breakup,Zhang2024Topological}, ultracold atomic arrays with tailored dissipation~\cite{Li2020TopologicalSwitch,zhou2022engineering,Liang2022Dynamic,Cai2024NHSE,zhao2025two}, and metamaterials~\cite{Rafi-Ul-Islam2022Interfacial,Wu2022NH2nd,Liu2023Experimental,ochkan2024non,zhang2025non,Jana2025Invisible,haydar-review-2025}. By providing a general prescription for flat-band engineering in open quantum matter, our results pave the systematic way for realizing interaction-driven and topological phases far from equilibrium.~\cite{jurivcic2024yukawa,Xiang-2024,murshed2024quantum,murshed2024yukawalorentz,PinoPRB-2025,yoshida2019non,Yamamoto2019Theory,Zhang2020Skin,Liu2020Mott,kawabata2019topological,Cayao2022Exceptional,Ghosh2022Non,Wang2022Emergent,Li2024Non,Yoshida2024Non,Ji2024Non,Shi2024Two-dimensional,Takemori2024Theory}.

\emph{General construction principle.} 
Flat bands naturally arise in Hermitian bipartite crystalline lattices (BCLs), where two interpenetrating sublattices, $L_A$ and $L_B$, host different numbers of degrees of freedom per unit cell, $N_A$ and $N_B$, respectively~\cite{cualuguaru2022general}. 
In momentum space, the Bloch Hamiltonian assumes a block form
\begin{equation}
H_\mathbf{k} = 
\begin{pmatrix}
A_\mathbf{k} & S_\mathbf{k} \\
S_\mathbf{k}^\dagger & B_\mathbf{k}
\end{pmatrix},
\end{equation}
with $A_\mathbf{k}$ and $B_\mathbf{k}$ describing intra-sublattice hoppings and $S_\mathbf{k}$ the inter-sublattice hybridization. 
If $A_\mathbf{k}$ hosts a momentum-independent eigenvalue $\epsilon_a$ with degeneracy $n_a > N_B$, the coupling $S_\mathbf{k}$ cannot fully hybridize these modes with $L_B$, leaving a kernel of dimension $(n_a - N_B)$ that manifests as dispersionless or flat bands at energy $\epsilon_a$, independent of $B_\mathbf{k}$. 

The main wisdom regarding the generalization of this principle to  the NH realm is provided by the chiral BCL Hamiltonian 
\begin{equation}
H^\mathcal{C}_\mathbf{k} = 
\begin{pmatrix}
0 & S_\mathbf{k} \\
S_\mathbf{k}^\dagger & 0
\end{pmatrix},
\label{eq:chiral-BCL}
\end{equation}
with at least $(N_A - N_B)$ flat bands~\cite{lieb1989two},  
possessing the chiral (sublattice or unitary particle-hole) symmetry, $\{\mathcal{C},H^\mathcal{C}_\mathbf{k}\}=0$, with the chiral operator 
\begin{equation}
\mathcal{C} = \begin{pmatrix}
\boldsymbol{1}_{N_{A}} & 0 \\
0 & -\boldsymbol{1}_{N_{B}}
\end{pmatrix},
\end{equation}
 where $\boldsymbol{1}_n$ is the unitary $n\times n$ matrix. Therefore, $(\mathcal{C} H^\mathcal{C}_\mathbf{k})^\dagger=-\mathcal{C} H^\mathcal{C}_\mathbf{k}$, implying that the operator $\mathcal{C} H^\mathcal{C}_\mathbf{k}$ is anti-Hermitian. {Importantly, this is representation–independent: for any unitary, Hermitian chiral operator \(\mathcal C\) with \(\mathcal C^2=1\), the constraint \(\{\mathcal C,H\}=0\) guarantees an off–diagonal (bipartite) form of the operator $H$ in the chiral (\(\mathcal C\)) basis. See also Sec. S1. of the Supplementary Material (SM)~\cite{SI}.} Consequently, we use this  operator to construct the   NH deformation of the chiral-symmetric Hermitian Hamiltonian $H^\mathcal{C}_\mathbf{k}$,
\begin{equation}\label{eq:NHham-master}
H_\mathbf{k} = (1+\alpha \mathcal{C}) H^\mathcal{C}_\mathbf{k} =
\begin{pmatrix}
0 & (1+\alpha) S_\mathbf{k} \\
(1-\alpha) S_\mathbf{k}^\dagger & 0
\end{pmatrix},
\end{equation}
where $\alpha \in \mathbb{R}$ quantifies the non-Hermiticity. Notice that the chiral operator $\mathcal{C}$ plays the role analogous to the mass term in the case of NH Dirac Hamiltonian~\cite{jurivcic2024yukawa} since it anticommutes with the parent Hermitian Hamiltonian, $H^\mathcal{C}_\mathbf{k}$. 
Such deformation leaves the original  flat bands intact [as shown in Sec.~S1 of SM~\cite{SI}] while introducing NH spectral features, as we show in the following.

To this end, we can express the eigenvalues and eigenstates of such Hamiltonian in terms of the singular eigenstates of $S_\mathbf{k}$ and $S^\dagger_\mathbf{k}$, denoted by $\psi_{n,\mathbf{k}}$ and $\phi_{n,\mathbf{k}}$, respectively, for $1\leq n\leq r_\mathbf{k}$, with $r_\mathbf{k}\leq N_B$ as the rank of $S_\mathbf{k}$, both with corresponding eigenvalue $\epsilon_{n,\mathbf{k}}\in\mathbb R$, and related as $S^\dagger_\mathbf{k}\phi_{n,\mathbf{k}}=\epsilon_{n,\mathbf{k}}\psi_{n,\mathbf{k}}$ and $S_\mathbf{k}\psi_{n,\mathbf{k}}=\epsilon_{n,\mathbf{k}}\phi_{n,\mathbf{k}}$. Then, the system have $2r_\mathbf{k}$ dispersive right-eigenstates of $H_\mathbf{k}$, given by
\begin{equation}
    \Psi^\pm_{n,\mathbf{k},R}=\frac{1}{\sqrt{2\sqrt{1-\alpha^2}}}\begin{pmatrix}
\pm\sqrt{1+\alpha}\,\phi_{n,\mathbf{k}} \\ \sqrt{1-\alpha}\,\psi_{n,\mathbf{k}}
    \end{pmatrix},
\label{eq:chiral_dispersive_states}
\end{equation}
with the dispersion  
\begin{equation}
E_{n,\mathbf{k}}^\pm = \pm \sqrt{1-\alpha^2}\,\epsilon_{n,\mathbf{k}},
\end{equation}
for $1\leq n\leq r_\mathbf{k}$.
The corresponding left-eigenstates can be obtained immediately by the substitution $\alpha\to-\alpha$, and the normalization factor is chosen to satisfy the biorthogonal condition $\Psi^\dagger_{\mathbf{k},L}\Psi_{\mathbf{k},R}=1$, see also Sec.~S1 of SM~\cite{SI}. At the EPs ($|\alpha|=1$), the dispersive branches therefore  coalesce producing { EFBs} with biorthogonal states. Furthermore, beyond these spectral singularities, for $|\alpha|>1$, these modes persist, remaining dispersionless {throughout the Brillouin zone} while acquiring finite lifetimes as their energies turn purely imaginary~\cite{remark-1}.

To extend this construction, we break the chiral symmetry via asymmetric chemical potentials $\mu_{A,B}=\bar{\mu}\pm \delta\mu$ modifying  the Hamiltonian to
\begin{equation}
H_\mathbf{k}=
\begin{pmatrix}
\mu_A \boldsymbol{1}_{N_A} & (1+\alpha) S_\mathbf{k} \\
(1-\alpha) S_\mathbf{k}^\dagger & \mu_B \boldsymbol{1}_{N_B}
\end{pmatrix},
\label{eq:non-chiral-const}
\end{equation}
with the spectrum { for dispersive states} $E^\pm_{n,\mathbf{k}}=\bar{\mu}\pm\Delta_\mathbf{k}$, where $\Delta_\mathbf{k}=\sqrt{\delta\mu^2+(1-\alpha^2)\epsilon_{n,\mathbf{k}}^2}$. {Furthermore, as shown in Sec.~S1 of the SM~\cite{SI}, the originally Hermitian} flat bands are shifted to $\mu_A$ or $\mu_B$ while preserving their degeneracy. {Thus, the construction principle remains operative: whenever one sublattice hosts a momentum-independent eigenvalue whose degeneracy exceeds the dimensionality of its complement, flat bands necessarily emerge.}

{By contrast with the degeneracy–mismatch mechanism above, the EFBs we now unveil require no sublattice imbalance: a momentum–independent term that anticommutes with chiral BCL Hamiltonian~\eqref{eq:chiral-BCL} suffices to generate additional EPs, for $|\alpha|=1$ and \(\Delta_{\mathbf k}=0\), and collapse dispersive states into dispersionless modes at energy $\mu_{A,B}$ and \(\bar\mu\), respectively.
Notably, at the momentum-dependent second-order EPs $\Delta_\mathbf{k}=0$ the collapse persists even when \(S_{\mathbf k}\) is full rank, yielding flat-band states supported on both sublattices, which represents a  hallmark of NH physics.
Away from such EPs (\(\Delta_{\mathbf k}\in\mathbb C\)), the formerly dispersive branches appear as complex-conjugate pairs with a \(k\)-independent real part and \(k\)-dependent decay/gain rates, \(\mathrm{Im}\,E_{\bf k}\equiv\pm|\Gamma_{\bf k}|\), realizing an effective balanced gain–loss population within the flat bands (Sec.~S1 of the SM~\cite{SI}).
This spectral-collapse mechanism follows solely from the anticommutation between the chiral NH BCL Hamiltonian [Eq.~\eqref{eq:chiral-BCL}] and the momentum-independent term \(\delta\mu\,\mathcal C\) in Eq.~\eqref{eq:non-chiral-const}, which plays the role of a Dirac mass–like operator (Sec.~S2 of the SM~\cite{SI}). Consequently, any momentum-independent deformation that anticommutes with Eq.~\eqref{eq:chiral-BCL} {and squares to a multiple of the identity} produces the same effect (see also Fig.~\ref{fig:non_chiral_tenband} for an explicit example).

\begin{figure}
    \centering
    \includegraphics[width=\linewidth]{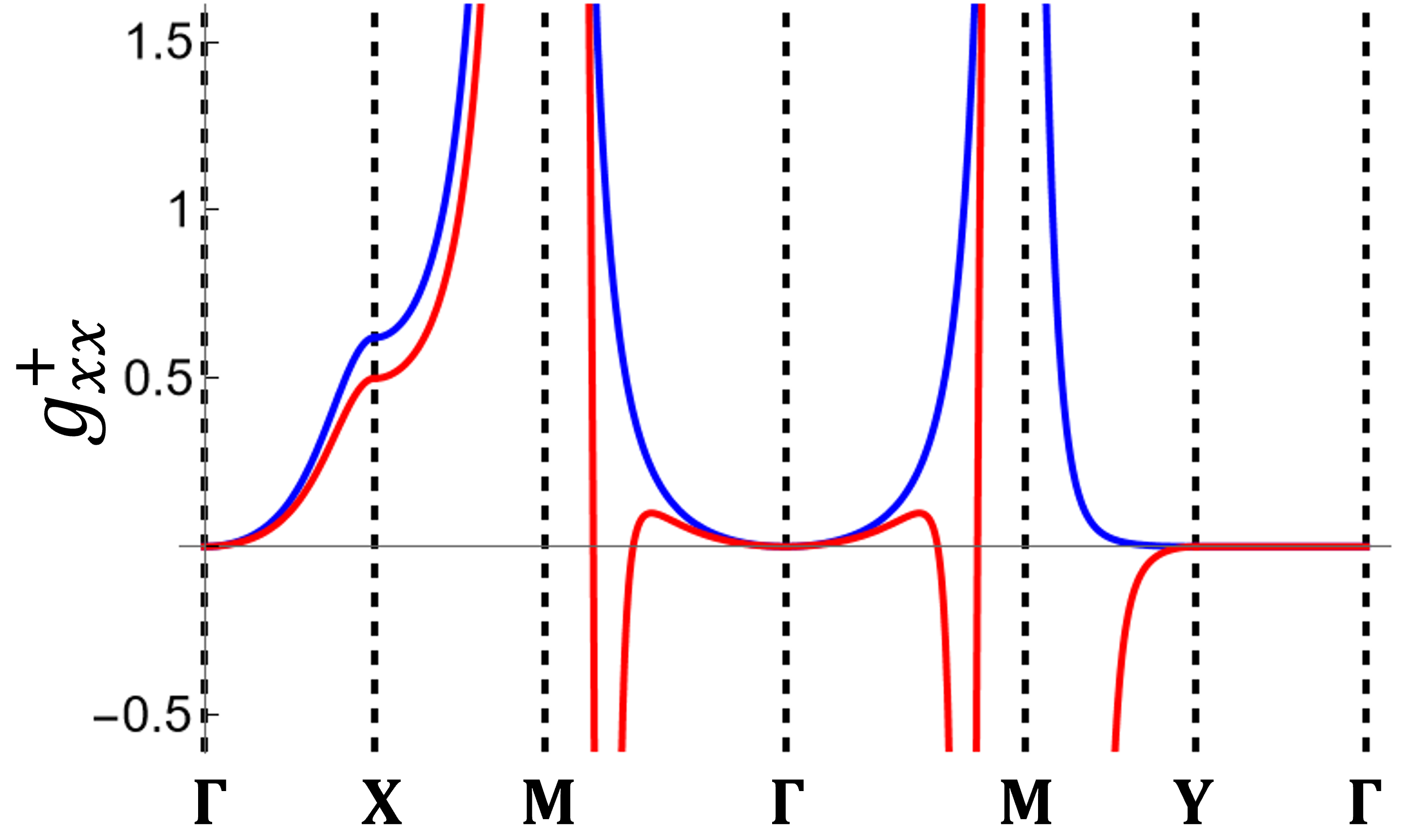}
    \caption{The diagonal element of the quantum metric in the Lieb lattice model with unbalanced chemical potential for $\alpha=0$ (blue) and $\alpha=1.1$ (red). See Eq.~\eqref{eq:QMetric}. In both cases, there is a divergence at the $M$-point in the Brillouin zone due to the standard band-touching, but the NH model exhibits an additional purely NH induced divergence to {negative} infinity, where $g^+_{xx}\to-\infty$, due to the spectral collapse onto the exceptional flat band pinned at $\bar\mu=0.5$. See Sec.~S4 of the SM for the details of the model. We here set $t=1$. See Fig.~S3 in the SM for the corresponding plots of the band structure corroborating the emergence of the EPs at finite momenta associated with the singularity in the quantum metric.}
    \label{fig:quantum_metric}
\end{figure}

\emph{Generalized NH BCL.}
We next explore the flat-band principle in its most general bipartite form, where a single sublattice hosts a momentum-independent eigenvalue while its counterpart carries arbitrary form~\cite{cualuguaru2022general}. The Hamiltonian can be written as
\begin{equation}
H_\mathbf{k} =
\begin{pmatrix}
\epsilon_a\,\boldsymbol{1}_{N_A} & (1+\alpha) S_\mathbf{k} \\
(1-\alpha) S_\mathbf{k}^\dagger & B_\mathbf{k}
\end{pmatrix},
\label{eq:Hgen}
\end{equation}
where $B_\mathbf{k}$ represents an arbitrary, momentum-dependent intra-sublattice operator. In this case, the flat bands associated with the kernel of $S_\mathbf{k}^\dagger$ remain pinned at $E_{n,\mathbf{k}} = \epsilon_a$, since their existence relies only on the degeneracy mismatch and the singular structure of $S_\mathbf{k}$, which are unaffected by the NH deformation. However, unlike in the previous cases, the presence of $B_\mathbf{k}$ introduces nontrivial momentum dependence into the dispersive sector. As a result, even at the EPs $\alpha = \pm 1$, the dispersive bands do not necessarily collapse onto the flat bands. The spectral flattening observed in simpler models is thus no longer guaranteed, as it depends on the detailed structure of $B_\mathbf{k}$.

\emph{Topological aspects.} To probe the geometric properties of the EFBs, we evaluate the NH quantum geometric tensor (QGT)~\cite{shen2018topological,zhu2021band,hu2024generalized},
\begin{equation}\label{eq:QMetric}
    Q^n_{\mu\nu} =
    \langle \partial_\mu \psi_n^L | \partial_\nu \psi_n^R \rangle 
    - 
    \langle \partial_\mu \psi_n^L | \psi_n^R \rangle
    \langle \psi_n^L | \partial_\nu \psi_n^R \rangle,
\end{equation}
whose real and imaginary parts define the quantum metric, $g^n_{\mu\nu} = \mathrm{Re}\, Q^n_{\mu\nu}$, and the Berry curvature, $F^n_{\mu\nu} = -2\,\mathrm{Im}\, Q^n_{\mu\nu}$, respectively.

In chiral NH BCLs, the deformation acts as a global prefactor without altering the momentum dependence of the eigenstates. Consequently, the Berry curvature vanishes, and the flat bands inherit the trivial topological character of their Hermitian counterparts~\cite{cualuguaru2022general,kruchkov2022quantum}. This behavior persists even in the presence of sublattice-asymmetric chemical potentials and at EPs: the left- and right-eigenstates remain identical to those of the Hermitian model, preventing the generation of new topological invariants. 

Nevertheless, the {quantum metric} exhibits a distinct and experimentally observable sensitivity to NH effects~\cite{LiaoPRL2021_metricNH,RenNatCommun2021_metric}. In our case, dispersive states in systems with finite chemical potential acquire a non-trivial $\alpha$-dependence in $g^n_{\mu\nu}$ (see Sec.~S3 of SM~\cite{SI}). Since the quantum metric governs measurable responses,  this dependence provides a concrete route to detecting NH signatures in flat-band systems. Thus, while the addition of NH terms does not alter the conventional topological classification, it leaves a geometric imprint encoded in the quantum metric, such as an additional divergence close to high-symmetry points in the Brillouin zone, as shown in the case of a  Lieb lattice model (see Sec.~S4 of the SM~\cite{SI}) in Fig.~\ref{fig:quantum_metric}. { We can understand this additional divergence as a hallmark of the {self-orthogonality} phenomenon at the EPs, that can be quantified through the blowup of the Petermann factor $K$, which describes the sensitivity of the NH system to perturbations~\cite{Petermann1979_SpontaneousEmissionFactor,Siegman1989_ExcessSpEmission_II,Wiersig2023_PetermannPhaseRigidity}, and, in turn,   has direct bearings  on the observables in their vicinity. Indeed, near an EP, the divergence of the Petermann factor drives the scaling of the quantum metric, \(g\!\sim\!K\), so the energy absorption is enhanced in their vicinity~\cite{Wiersig2023_PetermannPhaseRigidity,LiaoPRL2021_metricNH,EsinPRL2025}. Furthermore, in noncentrosymmetric magnetic crystals enjoying \(\mathcal{PT}\) symmetry, the metric-dipole-controlled intrinsic nonlinear conductivity~\cite{GaoScience2023_metricNLH,DasPRB2023_metric} should be parametrically enhanced in the clean limit, while the scattering-time-dependent Drude pieces of the nonlinear conductivity concomitantly suppressed~\cite{remark}. See Sec.~S5 of the SM~\cite{SI} for a detailed discussion.

\begin{figure}[t]
    \centering
    \includegraphics[width=\linewidth]{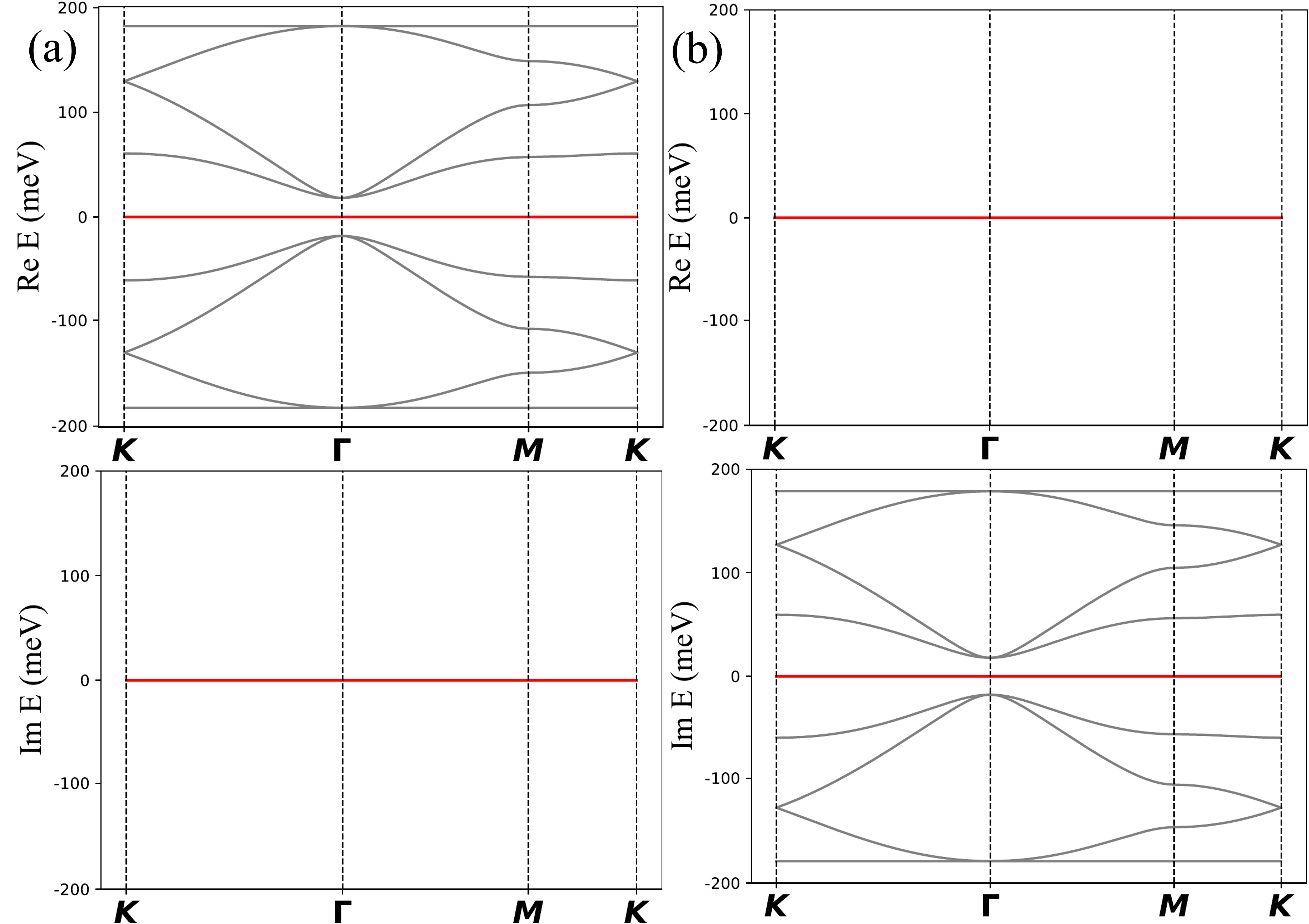}
    \caption{Real (top) and imaginary (bottom) part of the energy bands for the chiral ($\mu_A=\mu_B=0$) model for $\alpha=0$ (a) and $\alpha=1.4$ (b), with the same parameters employed in Ref.~\cite{po2019faithful}. In red, the degenerate flat band pinned at zero energy due to the difference between sublattices' sites, $N_A-N_B=2$. As elaborated in the main text, for this case the spectrum becomes purely imaginary beyond the EP $\alpha>1$, and thus all the states collapse onto the zero-energy level.}
    \label{fig:chiral_tenband}
\end{figure}

\emph{Examples.}
To illustrate how our construction operates in realistic settings, we apply it to two representative models. 

\begin{figure}[t!]
    \centering
    \includegraphics[width=\linewidth]{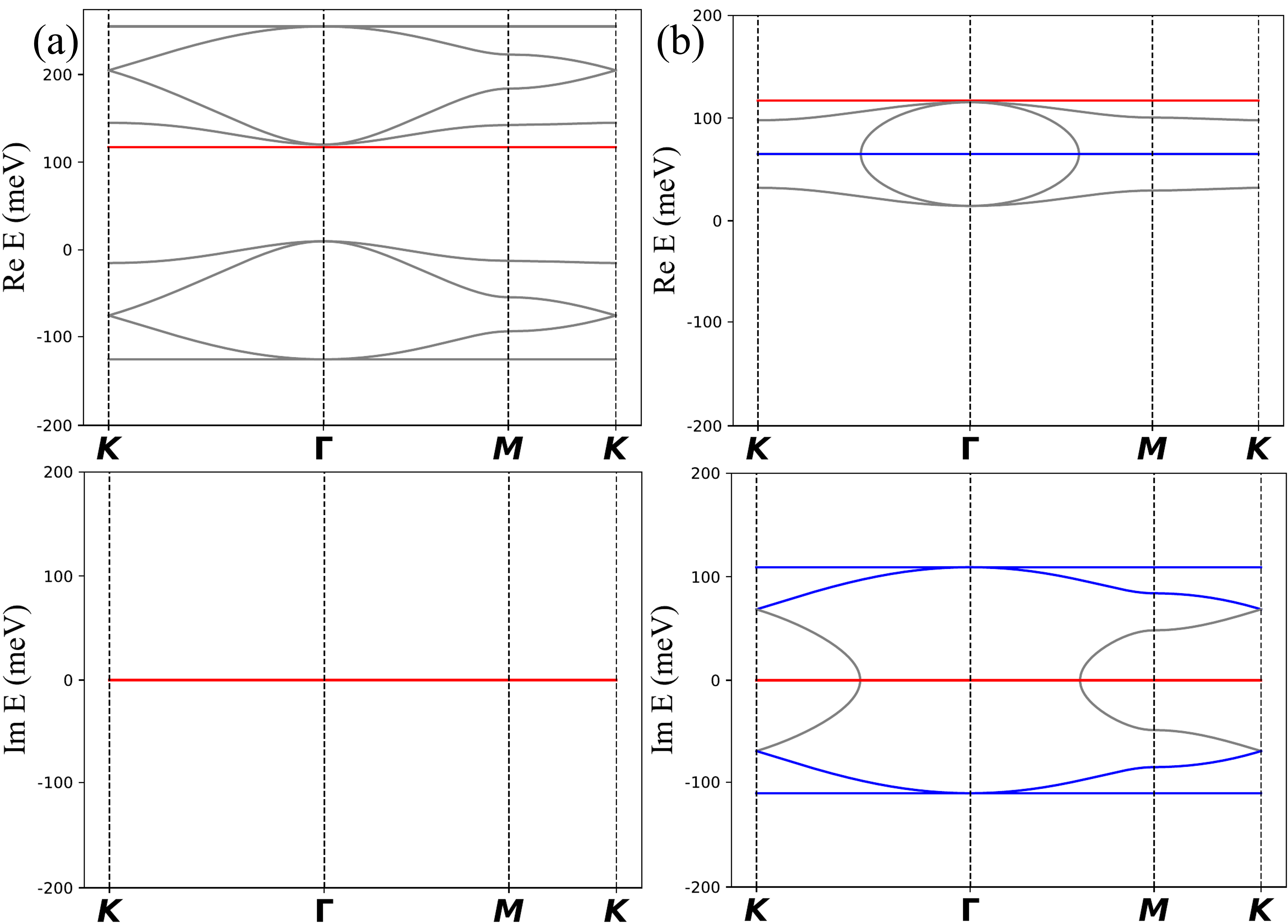}
    \caption{Real (top) and imaginary (bottom) parts of the energy bands for the non-chiral model for $\mu_A=117\,meV$ and $\mu_B=13\,meV$ with $\alpha=0$ (a) and $\alpha=1.2$ (b). Notice that the full energy spectrum is shifted, and in particular the flat band (in red) is pinned at $\mu_A$. Then, for $\alpha>1$, the dispersive states begin to collapse on to the value $\bar{\mu}=65\,meV<\mu_A$ {(in blue)} for an extended region of the BZ, {signaling the emergence of the exceptional flat bands}.}
    \label{fig:non_chiral_tenband}
\end{figure}

We first consider the ten-band model of TBG~\cite{po2019faithful}, by introducing nonreciprocal hopping between sublattices to realize a NH chiral BCL. As shown in Fig.~\ref{fig:chiral_tenband}, the spectrum evolves from being purely real to purely imaginary upon crossing the EP at $\alpha = 1$. For $\alpha < 1$, the flat bands persist with reduced dispersion, shrinking by a factor $\sqrt{1-\alpha^2} < 1$. When finite and unequal chemical potentials are applied to the sublattices, the flat bands shift in energy (Fig.~\ref{fig:non_chiral_tenband}). In this non-chiral configuration, the original flat band is pinned at $\mu_A$ for $\alpha < 1$, while tuning beyond the EP populates an additional { EFB} centered at $\bar{\mu}$. This emergent band is energetically favorable when $\bar{\mu} < \mu_A$, and although its states acquire a finite decay/gain rates $\tau\sim \pm 1/|{\rm Im}\, E|$,  (see Fig.~\ref{fig:non_chiral_tenband}), they get stabilized by becoming  long-lived at momentum-dependent EPs $|\epsilon_{n,\mathbf{k}}| = |\delta\mu/\sqrt{1-\alpha^2}|$, {for which $\tau\to\infty$ as $|\textrm{Im}E|\to0$}.

As a second example, we consider the NH extension of the tight-binding model for  Ca$_2$Ta$_2$O$_7$ in Ref.~\cite{cualuguaru2022general}, with its dispersion shown in Fig.~\ref{fig:Ca2Ta2O7}. Here, a large on-site energy imbalance ($\delta\mu \sim 5\,\mathrm{eV}$) ensures that the standard Hermitian flat band, pinned at $\mu_A < \bar{\mu}$, remains energetically dominant. Importantly, due to the substantial chemical potential offset, the spectrum remains real over a broad range of $\alpha > 1$, indicating that the system undergoes unitary evolution despite NH deformation. Consequently, the flat-band physics is only weakly modified, demonstrating robustness against strong on-site asymmetry.

These examples highlight the versatility of our framework: it captures NH spectral collapse and the emergence of EFBs in systems with tunable asymmetry, while also describing regimes where conventional flat bands remain largely unaffected by the NH effects.

\begin{figure}[t!]
    \centering
    \includegraphics[width=\linewidth]{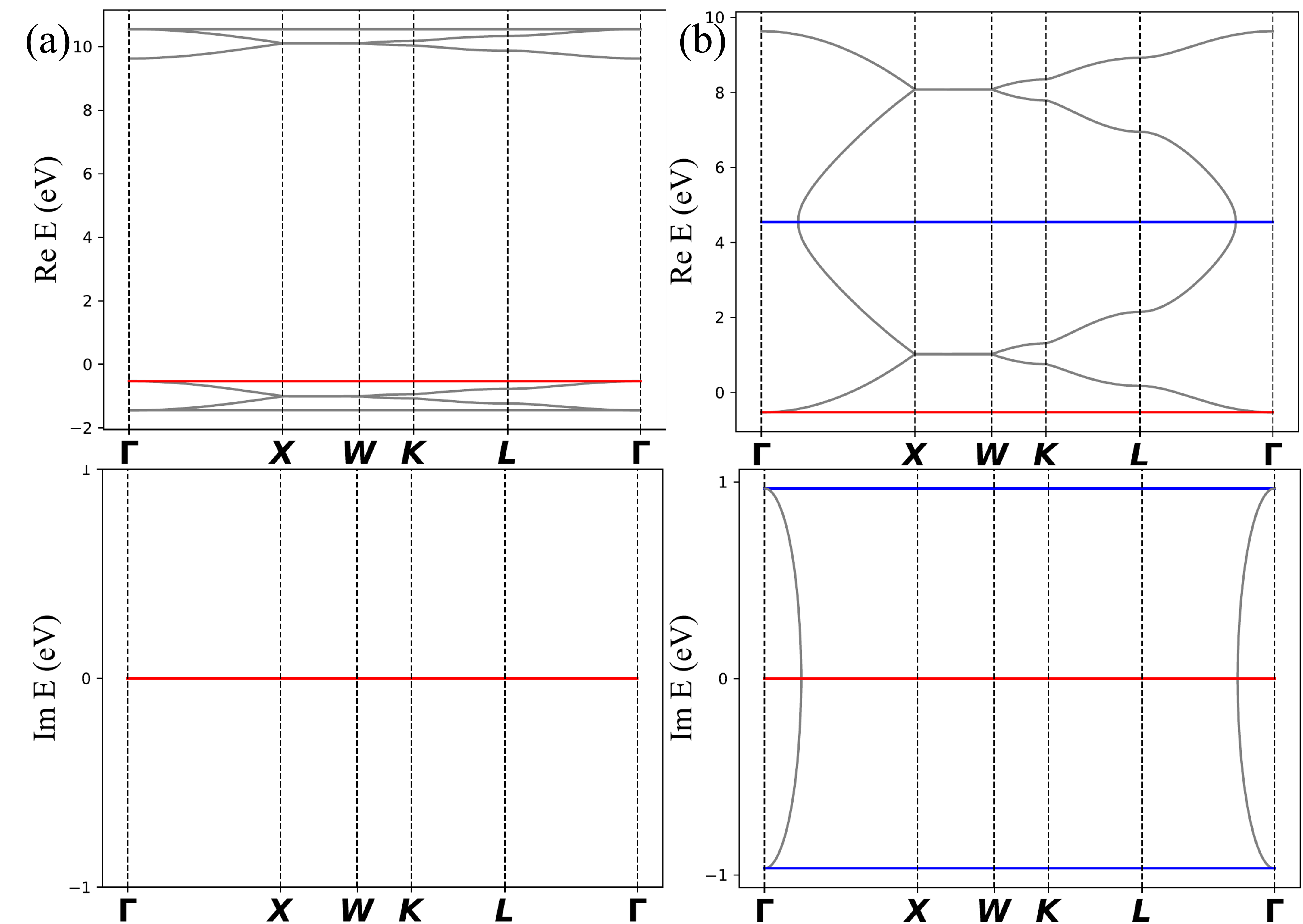}
    \caption{Energy spectrum of the tight-binding model for the Ca$_2$Ta$_2$O$_7$ compound, with the hopping parameters in Ref.~\cite{cualuguaru2022general}, for $\alpha=0$ (a) and $\alpha=1.9$ (b). For smaller values of $\alpha$, the spectrum is still real-valued, due to the large gap between on-site energies, $\delta\mu\approx 5$eV. Here again the Hermitian flat band pinned at $\mu_A\approx-0.5\,eV$ is highlighted in red, {and the exceptional flat bands are shown in blue}.}
    \label{fig:Ca2Ta2O7}
\end{figure}

\emph{Discussion and outlook.} We have shown that the flat-band mechanism, where a sublattice imbalance enforces dispersionless modes, carries over unchanged from the realm of Hermitian to NH crystals. In this broader setting, the mechanism not only reproduces the Hermitian flat bands but also generates a distinct class of EFBs via spectral collapse at EPs, where dispersive modes collapse to form the EFBs. Their energies and lifetimes are tunable, with their biorthogonal eigenstates spanning both sublattices, features absent in closed systems, as is the case of a bipartite lattice with imbalanced but constant sublattice chemical potentials.

{We point out that the NHSE~\cite{Foa-2018,yao-wang-PRL2018,LeeThomale2019_PRB_AnatomySkinModes} in our symmetry-based NH chiral BCL is absent by construction~\cite{Zhang2022Universal,Kawabata2019Symmetry,salib2023}. Even after breaking chiral symmetry, the spectrum remains with zero-area (see Fig.~S1 and  SM~\cite{SI}), thus NHSE is still absent~\cite{Zhang2022Universal}. By contrast, for the Hamiltonian~\eqref{eq:Hgen} with generic $B_{\mathbf k}$, skin modes are expected with model-dependent features. Furthermore, in lattices hosting Hermitian compact localized states~\cite{Flach_2014,Hwang2021General}, it is conceivable that the chiral-symmetric NH deformation (Eq.~\eqref{eq:NHham-master}) preserves compact support for the EFB modes  for \(|\alpha|\neq1\), while the dc electric fields should enhance their localization~\cite{Kolovsky2018Topological}. The detailed study of localization features of EFB states in BCLs is left for future research.}

{We next clarify the validity regime of the effective NH Bloch description for our nonreciprocal tight-binding models. {In a Markovian open-system setting, the Lindblad master equation \(\dot\rho=-i[H_{\rm sys},\rho]+\sum_\ell(L_\ell\rho L_\ell^\dagger-\tfrac12\{L_\ell^\dagger L_\ell,\rho\})\) defines the  effective NH  operator \(H_{\rm eff}=H_{\rm sys}-\tfrac{i}{2}\sum_\ell L_\ell^\dagger L_\ell\) \cite{LindbladCMP1976}.} For these models, the effective NH Bloch Hamiltonian \(H_{\rm eff}\) accurately captures features related to single-particle correlators, such as non-Bloch spectra, exceptional points, and NHSE~\cite{YaoWang2018_PRL,OkumaSaito2020_PRL,Helbig2020_NatPhys}, whereas noise-sensitive observables require the full Lindblad operator~\cite{LindbladCMP1976}. {In particular, Green's-function probes such as DOS and Kubo conductivities are governed by \(H_{\rm eff}\), as realized in engineered nonreciprocity via cascaded reservoirs~\cite{MetelmannClerk2015_PRX}. In momentum space one may use linear jump operators \(L_{\nu,\mathbf k}=\sqrt{\gamma_{\nu,\mathbf k}}\big(\tilde a_{\nu,\mathbf k}+e^{i\phi_{\nu,\mathbf k}}\tilde b_{\nu,\mathbf k}\big)\), where \(\tilde a_{\nu,\mathbf k}\) and \(\tilde b_{\nu,\mathbf k}\) denote mode operators in the singular-vector basis of \(S_{\mathbf k}=U_{\mathbf k}\Sigma_{\mathbf k}V_{\mathbf k}^\dagger\), yielding the required asymmetric $A-B$ hoppings; see Sec.~S6 of SM for details.} {In our construction, the Bloch Hamiltonian \(H_{\mathbf k}\) in Eq.~\eqref{eq:NHham-master} should be viewed as such an effective \(H_{\rm eff}(\mathbf k)\) governing the single-particle features of a nonreciprocal Lindblad realization.} This regime has been demonstrated experimentally in an ultracold-atom platform~\cite{Liang2022Dynamic}.
 }



Photonic crystals with engineered gain–loss, ultracold-atom arrays with controlled dissipation, and non-reciprocal electronic metamaterials provide natural platforms for the NH BCLs considered here. In these settings, EFBs are expected to manifest via enhanced density of states, characteristic transport anomalies, and interaction-driven responses, thereby opening routes to correlated phases without Hermitian analogues. More broadly, the interplay of interactions, topology, and EFBs promises distinctive many-body phenomena in open quantum matter. Our symmetry-based construction offers a practical framework for flat-band engineering in NH materials and enables systematic exploration of strongly correlated and topological phases far from equilibrium.

\emph{Acknowledgment.}  We are grateful to Bitan Roy for the critical reading of the manuscript. This work is supported by Fondecyt (Chile) Grant  No.   1230933 (V.J.). J.P.E. acknowledges support from Agencia Nacional de Investigación y Desarrollo (ANID) – Scholarship Program through the Doctorado Nacional Grant No. 2024-21240412, and Dirección de Postgrado UTFSM through the PIIC Grant No. 28/2025.

\bibliography{Bibliography}

\begin{thebibliography}{132}
\bibitem [{rem({\natexlab{a}})}]{remark-1}%
  \BibitemOpen
  \href@noop {} {}\bibinfo {howpublished} {More generally, for $H'_C(\mathbf{k})=\bigl(\begin{smallmatrix}0 & A \\ B & 0\end{smallmatrix}\bigr)$, $(H'_C)^2=\mathrm{diag}(AB,BA)$, so zero-energy flat bands occur when $\ker(AB)$ is nontrivial; $N_A\neq N_B$ then guarantees $|N_A-N_B|$ zero modes. In our construction, $|\alpha|=1$ yields $AB=BA=0$ throughout the Brillouin zone, producing the EP collapse and exceptional-flat-band regime. See Sec.~S1 of the SM~\cite{SI} for more details.} ({\natexlab{a}})\BibitemShut {NoStop}%
\bibitem [{rem({\natexlab{b}})}]{remark}%
  \BibitemOpen
  \href@noop {} {}\bibinfo {howpublished} {This \(\mathcal{PT}\)-symmetric setting is invoked only as an experimentally relevant example, and does not constitute a requirement of our chiral-symmetric exceptional-flat-band construction.} ({\natexlab{b}})\BibitemShut {NoStop}%
\end{thebibliography}
\end{document}